\documentclass[final,5p,times,compress,twocolumn]{elsarticle}

\usepackage{amssymb}
\usepackage{lipsum}
\usepackage{amsmath}
\usepackage{booktabs}
\usepackage{xcolor}
\usepackage{array}
\usepackage{bigints}
\usepackage{orcidlink}
\usepackage{hyperref}
\usepackage{bm}
\usepackage{enumitem}

\journal{Physics Letters B}

\usepackage{graphicx}
\usepackage{dcolumn}
\usepackage{hyperref}
\usepackage[utf8]{inputenc}
\usepackage{multirow}
\usepackage{soul}
\usepackage{float}
\usepackage{graphicx}
\usepackage{times}
\usepackage[normalem]{ulem}
\usepackage{color}
\usepackage{cellspace}
\usepackage{lipsum}
\usepackage{adjustbox} 
\usepackage{rotating}
\usepackage{tikz}
\usepackage{changepage}

\newcommand{\be}{\begin{equation}}
\newcommand{\ee}{\end{equation}}
\newcommand{\bea}{\begin{eqnarray}}
\newcommand{\eea}{\end{eqnarray}}

\DeclareUnicodeCharacter{2212}{-}
\usepackage[none]{hyphenat}
\usepackage{booktabs}
\usepackage{url}
\usepackage{array} 
\usepackage{amsmath} 
\usepackage{xcolor}
\usepackage{array}
\usepackage{bigints}
\usepackage{orcidlink}
\usepackage{hyperref}
\usepackage{bm}
\usepackage[normalem]{ulem}
\usepackage{enumitem}

\begin{document}

\begin{frontmatter}



\title{Inferring the Equation of State from Neutron Star Observables via Machine Learning}


\author[label1,label2]{N. K. Patra \href{https://orcid.org/0000-0003-0103-5590}{\includegraphics[scale=0.06]{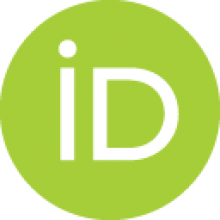}}}\ead{nareshkumarpatra3@gmail.com}

\author[label3]{Tuhin Malik\href{https://orcid.org/0000-0003-2633-5821}{\includegraphics[scale=0.06]{Orcid-ID.png}}}\ead{tm@uc.pt}

\author[label3]{Helena Pais\href{https://orcid.org/0000-0001-7247-1950}{\includegraphics[scale=0.06]{Orcid-ID.png}}}\ead{hpais@uc.pt}

\author[label1,label4]{Kai Zhou\href{https://orcid.org/0000-0001-9859-1758}{\includegraphics[scale=0.06]{Orcid-ID.png}}}\ead{zhoukai@cuhk.edu.cn }

\author[label5,label6]{B. K. Agrawal\href{https://orcid.org/0000-0001-5032-9435}{\includegraphics[scale=0.06]{Orcid-ID.png}}}\ead{bijay.agrawal@saha.ac.in}

\author[label3]{Constança Providência\href{https://orcid.org/0000-0001-6464-8023}{\includegraphics[scale=0.06]{Orcid-ID.png}}}\ead{cp@uc.pt}

\affiliation[label1]{School of Science and Engineering, The Chinese University of Hong Kong, Shenzhen (CUHKShenzhen), Guangdong, 518172, China}
\affiliation[label2]{Department of Physics, BITS-Pilani, K. K. Birla Goa Campus, Goa 403726, India.}

\affiliation[label3]{CFisUC, Department of Physics, University of Coimbra, P-3004 - 516  Coimbra, Portugal}

\affiliation[label4]{Frankfurt Institute for Advanced Studies, Ruth Moufang Strasse 1, D-60438, Frankfurt am Main, Germany}

\affiliation[label5]{Saha Institute of Nuclear Physics, 1 AF Bidhannagar, Kolkata 700064, India} \affiliation[label6]{Homi Bhabha National Institute, Anushakti Nagar, Mumbai 400094, India.}

 




\begin{abstract}
We have conducted an extensive study using a diverse set of equations of state (EoSs) to uncover strong relationships between neutron star (NS) observables and the underlying EoS parameters using symbolic regression method. These EoS models, derived from a mix of agnostic and physics-based approaches, considered neutron stars composed of nucleons, hyperons, and other exotic degrees of freedom in beta equilibrium. The maximum mass of a NS is found to be strongly correlated with the pressure and baryon density at an energy density of approximately 800 MeV.fm$^{-3}$.  We have also demonstrated that the EoS can be expressed as a function of radius and tidal deformability within the NS mass range 1-2$M_\odot$. These insights offer a promising and efficient framework to decode the dense matter EoS directly from the accurate knowledge of NS observables.
\end{abstract}


\end{frontmatter}




\section{Introduction} \label{introduction}
Neutron stars (NSs) are one of the most fascinating and mysterious objects in the universe \cite{Haensel2007, Glendenning:1997wn, Rezzolla:2018jee}. They provide a unique window into the physics of extremely dense matter. These remnants of supernova explosions are incredibly dense and compact, having mass upto about twice that of the Sun with the radii $\sim$ 10 km \cite{Nicholl:2020mkh, Piekarewicz2017, Ozel:2015fia, Woosley:2016hmi}.
The astrophysical observations of NS properties are crucial, as they offer a way to constrain the equation of state (EoS) of dense matter \cite{Lattimer:2012nd, Hebeler:2013nza, Das:2022ell, Lattimer:2021emm, Ferreira:2021pni, Zhang:2018vrx, Malik:2018zcf}. The EoS can be expressed as a relation between the pressure and energy density of NS matter. It is a critical input for understanding many astrophysical phenomena, such as supernovae and neutron star mergers \cite{LATTIMER2000121, Steiner:2012rk}. The time evolution of supernova gravitational waves strongly depends on the EoS, as demonstrated in Ref.~\cite{Sotani:2021ygu}. Recent advances in observational technology have provided new insights into NS properties, driving intense scientific interest and activity \cite{Abbot2017, Abbot2018, Abbot2019, Miller:2019cac, Riley:2019yda, Miller:2021qha, Riley:2021pdl}. Recycled millisecond pulsars undergoing starquakes show a sudden increase in gravitational wave amplitude, providing a unique signature of NS matter at high densities\cite{LIGOScientific:2025kei, Chatterjee:2024eje, Chatterjee:2021mex, Melatos:2005ez}.
These breakthroughs have inspired innovative research spanning multiple disciplines, leveraging collaborative and interdisciplinary approaches development. In particular, the intricate direct connections between EoS parameters and NS observables have become a focal point of study. Sophisticated computational techniques, such as machine learning, are increasingly employed to uncover hidden correlations within complex datasets, enabling researchers to explore relationships among EoS and NS properties with unprecedented precision \cite{Krastev:2021reh, Krastev:2023fnh}.

A number of machine learning (ML) techniques, particularly neural networks, have been applied in recent years to unravel the physics of neutron stars \cite{Cuoco:2020ogp, Whittaker:2022pkd, Ferreira:2021pni, Ferreira:2022nwh}. An analysis of simulated observations of the radius and tidal deformability of NSs has been used to determine the internal composition of NSs using Bayesian Neural Networks (BNNs) \cite{Carvalho:2023ele}. Using these models, the dense matter equation of state has also been inferred from various astrophysical and gravitational wave measurements \cite{Thete:2022eif}. In a model-independent approach, ML algorithms have been used to predict the binding energy of atomic nuclei from AME2016 datasets, yielding an outer crust EoS for nuclear matter \cite{Anil:2020lch}. The equation of state and non-monotonic behaviour of speed of sound has also been studied using ML in NS \cite{Chatterjee:2023ecc}. As well as traditional machine learning models, feed-forward artificial neural networks were used to extrapolate the separation energy of hypernuclei \cite{Vidana:2022prf}. It has also been shown that neural networks can handle complex non-linear relationships between observables and the underlying physics with respect to dense matter EoS reconstructions based on mass-radius (M-R) observations and other parameters of NSs \cite{Soma:2022qnv, Soma:2023rmq}. By utilizing the non-linear potential of neural networks, this innovative approach creates intricate mappings between observations (input) and the EoS table (output) \cite{Morawski:2020izm}. As a result of its application to NS physics, ML has not only provided insights \cite{Zhou:2023pti, He:2023zin} on par with or even exceeding traditional physics-based models but has also demonstrated a significant potential to reduce computational costs.

In addition to these methods, symbolic regression can provide a powerful alternative to uncover hidden correlations among different features in the datasets. Symbolic regression method (SRM), in particular, has emerged as a potent technique for identifying human-readable equations or relations hidden within datasets \cite{Bomarito2021, Zhang2023, Keren2023, Imam:2024xqg}. By utilizing the symbolic regression method, scientists can formulate comprehensible mathematical relationships, enhancing our understanding of the underlying physics of dense neutron star matter. Recent studies in physics have effectively leveraged symbolic regression techniques to derive analytical formulas and uncover underlying mathematical expressions from complex data \cite{Wadekar:2020oov}. From deriving particle kinematics in collider phenomenology to improving success rates in finding symbolic expressions related to physics problems, the symbolic regression method has demonstrated its power and adaptability \cite{Dong:2022trn}. Additionally, it has been instrumental in modelling astrophysical phenomena such as assembly bias, as well as rediscovering known physical laws such as the Gell-Mann–Okubo formula \cite{Zhang:2022uqk}. These applications exemplify the increasing collaboration between physics and machine learning and the great potential of the symbolic regression method to advance our understanding of the universe \cite{Udrescu:2019mnk, CAMELS:2020cof}. 

This paper investigates the potential of machine learning, particularly the symbolic regression method, to uncover the hidden relationships between neutron star properties and the equation of state.
For that purpose, we use a variety of neutron star equations of state, encompassing different compositions and several modelling frameworks, including meta-models. The EoS collection includes non-linear (NL) nucleonic and hyperonic (NL-hyp) models, models taken from the CompOSE database (CompOSE), models constructed from the extrapolation of the speed of sound (CSE), piecewise-polytropic (PWP) models, density-dependent models obtained from Bayesian inference (DDB), and models from a spectral representation (SR), providing a comprehensive foundation for our investigation. Exploring the relationships between various NS properties through correlation analysis offers a compelling approach to understand the dense matter EoS \cite{Carlson:2022nfb, Tews:2016jhi, Vidana:2009is, Margueron:2017eqc, Essick:2021kjb, Lim:2019som, Richter:2023zec, Patra:2022yqc, Patra:2023jbz, Patra:2023jvv, Imam:2023ngm, Sun:2024nye, Raithel:2016bux}.
Specifically, in this work, we applied a symbolic regression method to a wide-ranging dataset to investigate: (i) the relation between the NS maximum mass and EoS properties at different energy densities, and (ii) the interplay between neutron star properties and the central energy density, and pressure for the neutron star masses in the range of 1-2$M_\odot$.

The paper is organized as follows. In Section \ref{formalism}, we provide a concise overview of the formalism used in the present study. The comprehensive analysis of the results and the subsequent discussions are presented in Section \ref{results}. Finally, we provide a brief summary of the results in Section \ref{summary}.

\section{Methodology} \label{formalism}

\subsection{Collections of EoSs}\label{dataset}

We have used seven different sets of diverse EoSs for the present investigation as follows: 
\begin{itemize}
    \item NL: The EoS for $\beta$-equilibrium nucleonic matter with electrons and muons, obtained 
    within the relativistic mean-field (RMF) framework with non-linear interactions \cite{Malik:2023mnx}. To get the model parameters and EoSs, we use the constraints of pure neutron matter of EoS at low density from N$^3$LO calculation in chiral effective field theory \cite{Hebeler:2013nza, Lattimer:2021emm}, some nuclear saturation properties and the maximum mass of a neutron star greater than 2 $M_\odot$. 
    \item NL-hyp: Similar to NL but including hyperonic particles \cite{Malik:2022jqc}. We employ the same constraints as mentioned in the NL case to obtain the EoSs. We limit our consideration of hyperon couplings to a certain subset of couplings and use vector meson predictions from the SU(6) quark model. 
    \item CompOSE: {This set includes several neutron star matter EoSs from the CompOSE database \cite{CompOSE, CompOSECoreTeam:2022ddl, Typel:2013rza, Dexheimer:2022qhn}, including relativistic and non-relativistic models, with different compositions}. 
    \item CSE: Publicly available constructed EoSs using agnostic approaches within the inference frame subjected to various nuclear, astrophysical constraints and heavy-ion collision constraints, including speed of sound extrapolation models (CSE) \cite{Huth:2021bsp}. The EoSs at low density up to 1.5$\rho_0$ is constructed using chiral EFT input \cite{Hebeler:2013nza}. Above it, the EoSs are constructed using extrapolation in the speed of sound ({\color{blue}$c_s$}) in neutron star matter \cite{Tews:2018iwm}. This extrapolation is constrained solely by causality ({\color{blue}$c_s \leq c$}) and the stability of neutron-star matter ({\color{blue}$c_s \geq 0$}). 
    \item PWP: Obtained with similar constraints for the case of CSE but with Piecewise-polytrope (PWP) extension models  \cite{Huth:2021bsp}. Above density 1.5$\rho_0$, the EoSs are constructed using five polytropic segments with randomly chosen transition densities and polytropic indices \cite{Read:2008iy, Ozel:2009da}.
    \item DDB: The EoSs are constructed within the RMF framework with density-dependent coupling, constrained by existing observational, theoretical, and experimental data through Bayesian analysis \cite{Malik:2022zol}. The pure neutron matter constraints at low densities, derived from next-to-next-to-next-to-leading order (N$^3$LO) calculations in $\chi$EFT, are also applied in the construction \cite{Hebeler:2013nza, Lattimer:2021emm}. This EoS set is consistent with the NL dataset. 
    \item {Spectral representation (SR)}: This EoS set is obtained using another meta-model approach, namely the Spectral representation of EoS set \cite{Roy:2023gzi, Lindblom:2022mkr}. The methodology by Lindblom \cite{Lindblom:2010bb} leads to the smooth construction of the EoS for nuclear matter inside the NS by constructing spectral expansions of key thermodynamic quantities.
    \item All: Consist of all the above seven sets of EoSs. Due to the limited number of EoSs in the CompOSE set, we randomly sampled 100 EoSs from each set. 
\end{itemize} 

We use the above collection of EoSs to calculate several properties of neutron stars by using the Tolman-Oppenheimer-Volkoff (TOV) equations. We build a database containing these neutron star properties like mass (M), radius (R), and tidal deformability ($\Lambda$). Then the relation between these macroscopic properties and the EoS quantities such as energy density, and pressure, are explored,
where the mapping is obtained by applying a symbolic regression method.

\begin{figure}[!htb]
\centering
\includegraphics[width=1\linewidth]{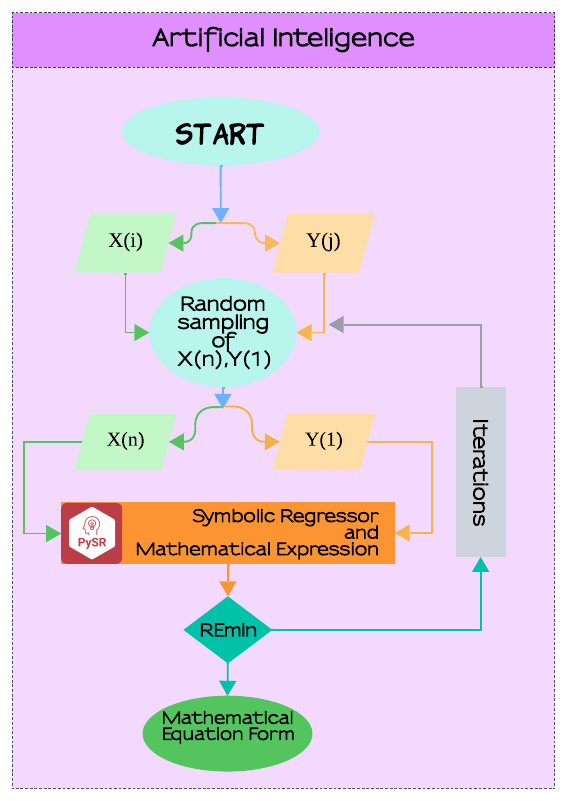}
\caption{Flowchart depicting the symbolic regression network used in this study (see the  text for details).}\label{fig1}
\end{figure}

\subsection{Sampling for symbolic regression}\label{sampling}

A number of advances have been made in machine learning, including techniques such as the symbolic regression method. Among these techniques, Python Symbolic Regression (PySR) stands out as a prominent algorithm employed for symbolic regression tasks. In addition to uncovering relationships within datasets, PySR conveys these relationships through mathematical equations that are human-readable. Although alternative machine learning methods or deep neural networks might yield superior predictive performance, their outcomes tend to be more complex in decoding the relations among various quantities. 
We applied the PySR algorithm as described in Refs.~\cite{cranmer2023, Tsoi_2024, Varun2011, Ferreira2019, Kumar:2024fui} to establish the relationship between various neutron star properties and the central density, energy density and pressure.

{Our objective is to derive a mapping from the thermodynamic variables of an equation of state (EoS)—such as energy density and pressure over various number densities—to the mass-radius-tidal deformability sequence of neutron stars. We also aim to correlate the maximum neutron star mass with these thermodynamic quantities. Because the maximum mass is a crucial constraint for a valid EoS and typically requires solving the computationally intensive TOV equations, establishing a semi-universal relation that predicts this mass without directly solving TOV could greatly reduce CPU usage in large-scale EoS inference. To achieve this, we will apply symbolic regression to our comprehensive EoS dataset, enhancing both efficiency and precision in our mapping.}

{ A comprehensive outline of our symbolic regression methodology is presented in the Figure~\ref{fig1}.  To initiate the process, we extract two types of vectors from our dataset: feature vectors
	$\bf{X(i)}$ and target vector $\bf{Y(j)}$. In these vectors, $i$ and $j$ represents the total number of feature and target quantity, respectively. We then enter the sampling phase, where we randomly select a subset of n feature ($\bf{X(n)}$) from the original characteristics ($\bf{X(i)}$). Along with these, we also choose a single random target $\bf{Y(1)}$ from the original set of targets $\bf{Y(j)}$. These randomly sampled combinations of features $\bf{X(n)}$ and target $\bf{Y(1)}$ are fed into the PySR algorithm to discover mathematical relationships between them. For effective hyperparameter tuning of the PySR algorithm, including parameters like the number of iterations and operators, we employ two optimization strategies: (i) Bayesian optimization and (ii) Grid search. The optimal equation is selected using a dual approach involving both techniques. The equation with the highest Pearson correlation coefficient and smallest relative error (RE in \%) between $\bf{X}$ and $\bf{Y}$ is identified as the best equation. This entire process is iterated over $\sim$ 1/2 million times to explore different feature combinations and identify the most robust mathematical expressions that relate our EoS quantities to the NS properties. We performed this analysis in two directions: first, using EoS quantities as features and NS properties as targets, and then reversed the approach by using NS properties as features and EoS quantities as targets to explore bidirectional relationships in the data.}

\section{Results and Discussion}\label{results}

\begin{figure*}[!htb]
    \centering
    \includegraphics[width=0.45\linewidth]{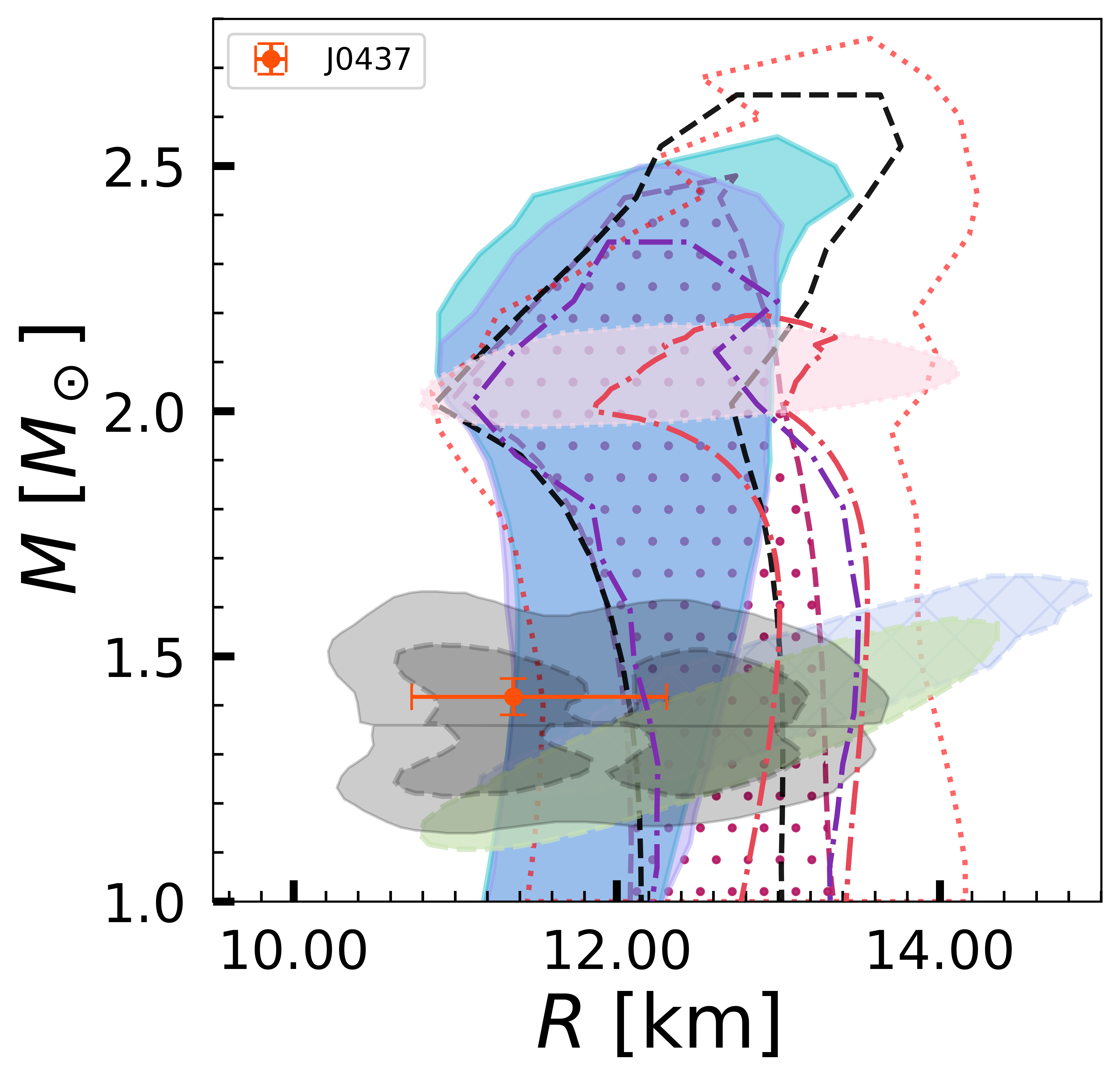}
    \includegraphics[width=0.45\linewidth]{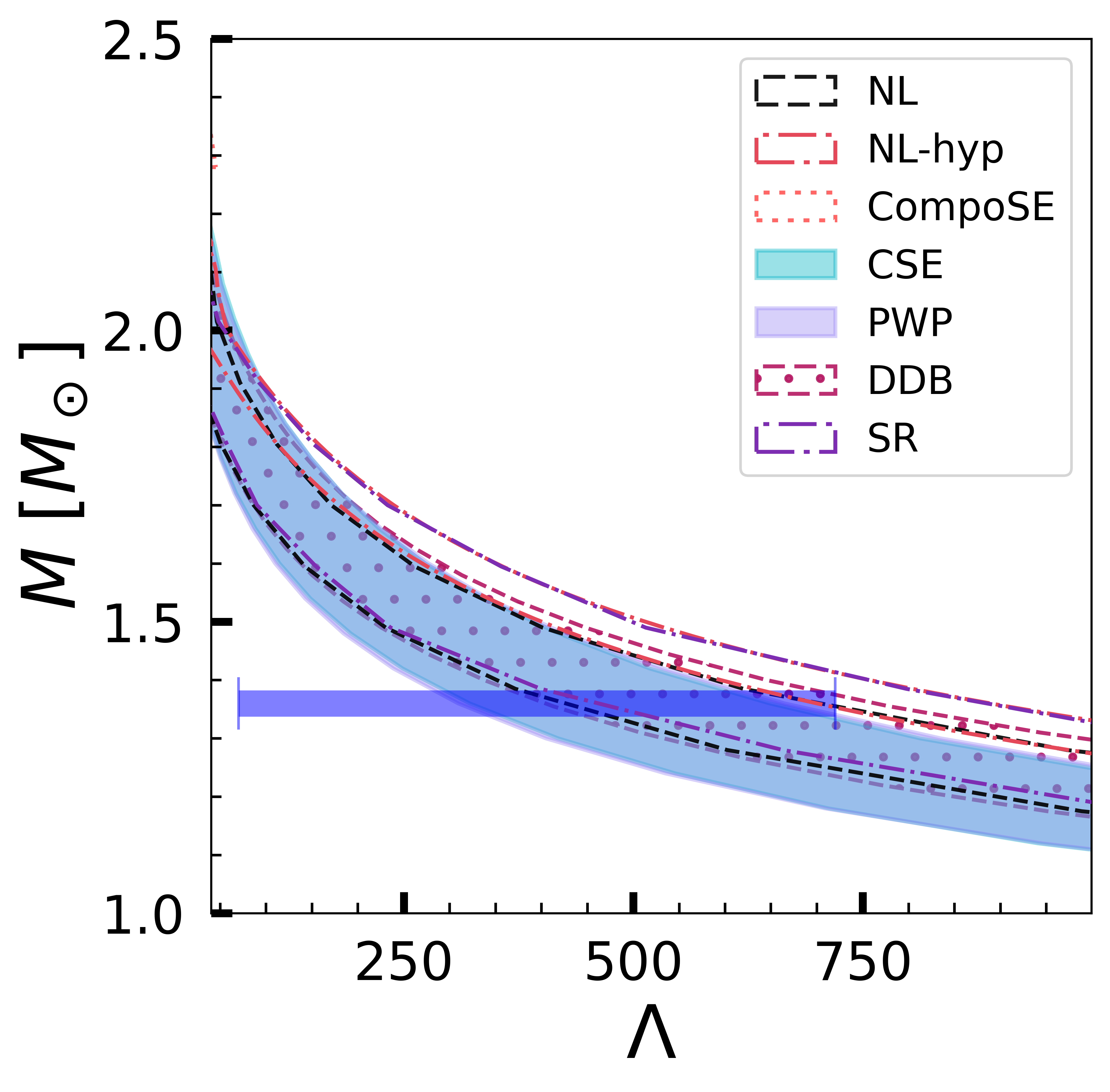}
    \caption{\textit{Left panel:} NS mass-radius regions for 90\% confidence intervals (CI) of the conditional probabilities $P(M|R)$ across seven datasets, shown with different colours and lines.  The grey shaded regions represent the 90\% (solid line) and  50\% (dashed line) CI of LIGO/Virgo constraints derived from the binary components of GW170817 event \cite{Abbott2018b}. The 1$\sigma$ (68\%) CI for the 2D posterior distribution in the mass-radius domain for the millisecond pulsar PSR J0030+0451 (in pastel blue and soft green) \cite{Riley:2019yda, Miller:2019cac} and PSR J0740+6620 (in blush pink) \cite{Riley:2021pdl, Miller:2021qha} from NICER X-ray data are also shown, together with the latest NICER measurements of mass and radius for PSR J0437-4715 \cite{Choudhury:2024xbk} (orange point). \textit{Right panel:} 
    Same as the left panel but for the mass-tidal deformability relation. The blue bar represents the constraints of the tidal deformability for the  $1.36M_\odot$ NS   \cite{Abbot2017}. } \label{fig2}
\end{figure*}

As previously discussed in Section \ref{dataset}, we use a diverse set of EoSs for cold dense neutron star matter, encompassing different compositions and several models including meta-models. These collections of EoS correspond to NL, NL-hyp, CompOSE, CSE, PWP, DDB, and SR models. Before embracing on our results, it is essential to examine the diversity of our EoSs concerning the neutron star properties. In Figure~\ref{fig2}, on the left, we display the 90\% confidence interval (CI) for the radius of neutron stars across masses from 1$M_\odot$ to the maximum mass obtained for each of the EoS set. This is compared with a range of astrophysical data. The shaded grey regions show the 90\% (solid line ) and 50\% (dashed line) confidence intervals of the LIGO / {\color{blue}Virgo} constraints, derived from the binary components of GW170817 \cite{Abbott2018b}. Moreover, the 1$\sigma$ (68\%) confidence intervals for the 2D posterior distribution in the mass-radius region for the millisecond pulsar PSR J0030+0451 are depicted in pastel blue and soft green \cite{Riley:2019yda, Miller:2019cac}. For the pulsar PSR J0740+6620, the confidence intervals, based on NICER X-ray data, are shown in blush pink. The most recent mass-radius data for the closest pulsar, PSR J0437-4715, is also included (orange) \cite{Choudhury:2024xbk}. The figure demonstrates that all our EoS collections are in harmony with the constraints imposed by astrophysical observations. {The CompOSE set covers the widest range in mass and radius as it contains EoS with different compositions, and models, relativistic and non-relativistic, {and most of the EOS were constrained using properties of finite nuclei, only sensitive to densities close to saturation density}.} On the right, the figure shows the mass against the dimensionless tidal deformability for all EoS collections. The blue bar indicates the tidal deformability constraints for the NS with mass 1.36$M_\odot$, as reported by Abbot {et al.}\cite{Abbot2017}. {The figure clearly shows that the various EoS sets examined in this study encompass a wide range of NS properties, whether in terms of mass-radius or mass-tidal deformability, all while adhering to astrophysical observation constraints.}

\subsection{ Direct Relation of NS Observables with EoS }
We now consider the relationships among EoS quantities and NS properties such as the maximum mass, radius and tidal deformability over a wide range of mass. We first explore the possibility of expressing the maximum NS mass in terms of pressure and baryon density at a suitable chosen value of energy density. Next, we express the central energy density and pressure as a function of NS mass, radius and tidal deformability.

\subsubsection{ Maximum NS Mass} \label{mapping_mmax}

\begin{figure*}[!ht]
    \centering
    \includegraphics[width=0.465\linewidth]{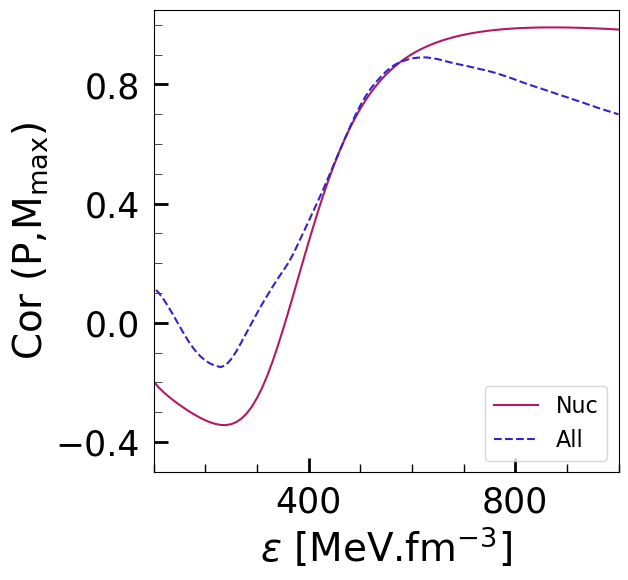}
    \includegraphics[width=0.45\linewidth]{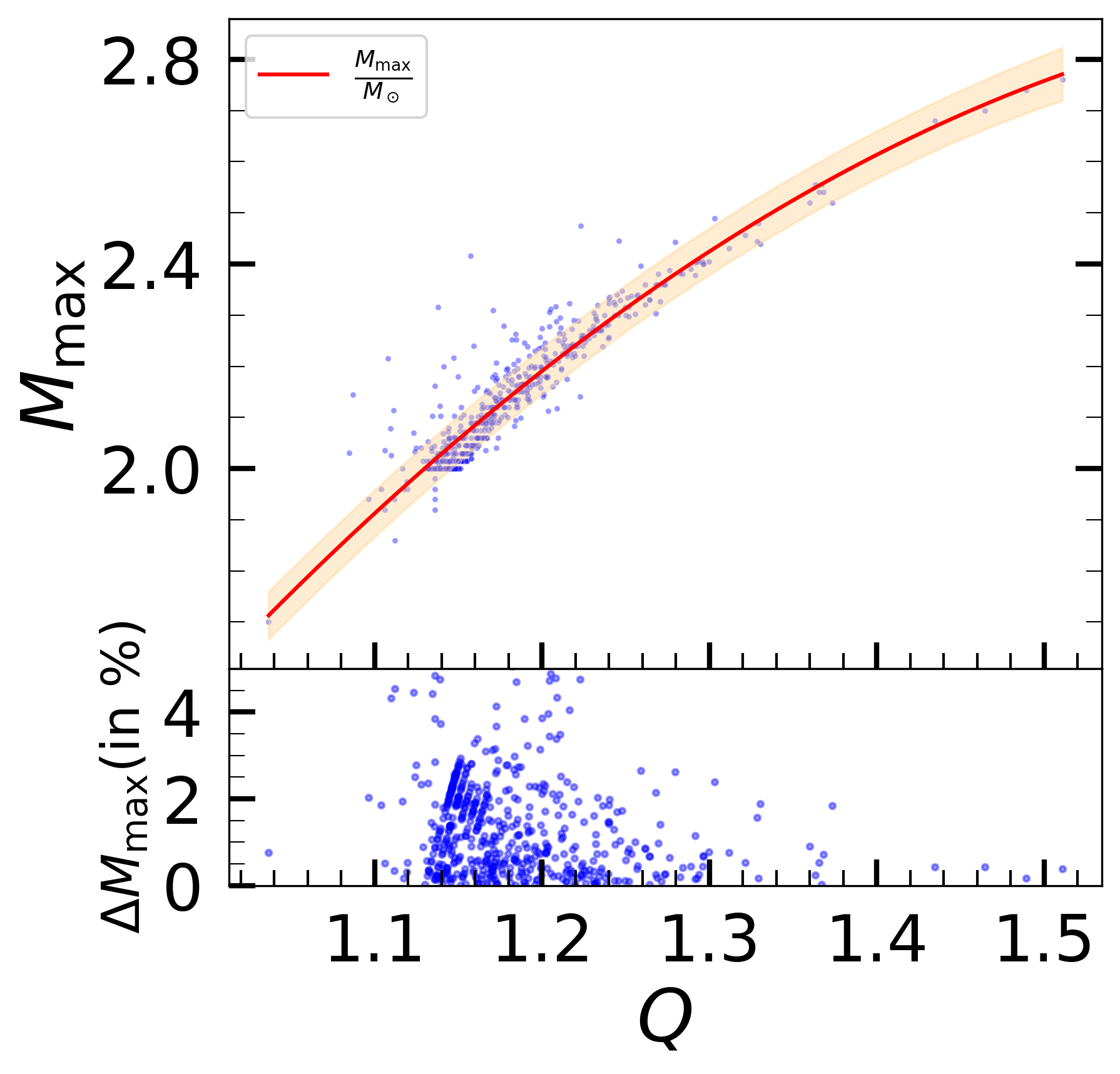}
    \caption{\textit{Left panel:} The values of Pearson's correlation coefficient obtained with pressure \( P \) and NS maximum mass ($M_{max}$) as a function of energy density. The purple and blue lines correspond to the results obtained for nucleonic EoSs (NL and DDB) labelled as "Nuc", and for the combined results of all seven datasets "All", respectively. \textit{Right panel:} The 90\% CI of the maximum mass of NS as a function of $Q$, given by Eq.~(\ref{q-fit}), for the nucleonic EoS. The red line is the quadratic fit, given by Eq.~(\ref{mmax}). The scatter points depict the results obtained by combining all sets of EoSs (in blue). The lower panel shows the relative error (in \%) for our fit (Eq.~(\ref{mmax})). }
    \label{fig3}
\end{figure*}

We aim to determine the relationship between the NS maximum mass and EoS properties such as energy density and pressure across different baryon densities in a model-independent manner using our diverse sets of EoS. 
In the left panel of Figure~\ref{fig3}, we display the values of the Pearson's correlation coefficient between pressure \( P \) and maximum mass of NS ($M_{max}$) as a function of energy density, for the nucleonic EoSs (purple line) that correspond to the NL and DDB datasets, and compare with those obtained for all the EoSs (blue dashed line).  We observe that the correlation becomes stronger within a specific range of energy densities for nucleonic EoSs. 
{This correlation is weak and negative at low energy densities and then increases up to ($r \sim 0.9$) at 800 MeV fm$^{-3}$ and tends to saturate. {For the combined EoS results (All), the correlation peaks at 600 MeV·fm\(^{-3}\) with a Pearson coefficient of 0.85. In contrast, the individual sets exhibit correlations \({\rm Cor}(P, M_{\rm max})\) of [0.99, 0.97, 0.94, 0.88, 0.97, 0.64, 0.75] for the DDB, NL, NL-hyp, SR, CompOSE, CSE, and PWP models, respectively. Beyond 600 MeV·fm\(^{-3}\), the overall correlation declines, reaching 0.8 at 800 MeV·fm\(^{-3}\), while the individual correlations adjust to [0.99, 0.94, 0.87, 0.78, 0.95, 0.37, 0.66] for the same models.} {As might be expected, the agnostic descriptions show the smallest correlations, with PWP models still showing a correlation above 70\%, 75\% at  $\epsilon=600$ MeV·fm\(^{-3}\). At $\epsilon=800$ MeV·fm\(^{-3}\), the \({\rm Cor}(P, M_{\rm max})\)  correlation suffers an overall decrease which is particularly strong for the agnostic descriptions: for CSE it goes  below 50\%.  }
Given these findings, we applied the symbolic regression method, as described in Sec.~\ref{sampling}, to further investigate potential relationships between NS maximum mass with pressure and baryon density at a given energy density. This analysis helped us establish a relationship for mixed EoS models at an energy density of 800 MeV fm\(^{-3}\).
The obtained equation is 
\bea
 \frac{M_{\rm max}}{M_\odot} = -4.0946 + 7.9151 Q - 2.2312 Q^2 ,\label{mmax} 
 \eea
 with
 \bea
 Q = 1.23326\times 10^{-6} \left(\frac{ P}{\rm MeV. \, fm^{-3}} + 0.77003\right)\frac{fm^{-3}}{\rho} \label{q-fit} \, , 
\eea
where \( P \) represents the pressure and \( \rho \) is the baryon density at an energy density of 800 MeV.fm\(^{-3}\). {In this analysis, we considered a total of 700 EoSs, where 100 EoSs were randomly selected from each of the seven different EoS sets. At a fixed energy density of 800 MeV/fm$^{3}$, we examined the corresponding baryon density $\rho$ values for all these EoSs. We found that the baryon density spans a range from 0.63 fm$^{-3}$ (minimum) to 0.77 fm$^{-3}$ (maximum), indicating noticeable variation across the different models in this high-density regime.} 
This correlation is shown in the right panel of Figure~\ref{fig3}. Overall, it illustrates that over 90\% of the models within the mixed set adhere closely to the derived equation, highlighting its robustness and practical applicability. The lower panel displays the fit error ($\Delta M_{\rm max}$) as a function of $Q$, which remains below 4\%, indicating the reliability of our fitted function. This relationship, being largely model-independent and reliable, will be employed for realistic applications in the following.

\subsubsection*{Applications}

\begin{table*}
\centering
\caption{The number of likelihoods, number of samples and the corresponding time, and CPU hours used for the two cases {considering set NL}, using the Pymultinest Sampler, that employs 1500 live points on the Deucalion HPC cluster. The system's large-x86 partition utilizes 4 nodes, each equipped with 128 CPUs. {For NL (w TOV) the TOV equations were solved during the sampling, for NL (w/o TOV)   Eq.~(\ref{mmax}) was applied to impose a maximum mass above two solar masses.}}  \label{tab1}
\setlength{\tabcolsep}{9pt}
\renewcommand{\arraystretch}{1.4}
\begin{tabular}{>{\raggedright\arraybackslash}m{2.5cm}>{\centering\arraybackslash}m{3cm}>{\centering\arraybackslash}m{3cm}>{\centering\arraybackslash}m{3cm}>{\centering\arraybackslash}m{3cm}}
\hline \hline
Model & Total Likelihood evaluated & Total Samples in Posterior & Sampling Time (Sec) & CPU Hr \\
\hline
NL (w TOV) & 307018 & 8291 & 6828.08 & 971.10 \\
NL (w/o TOV) & 283718 & 8245 & 896.49 & 127.50 \\
\hline \hline
\end{tabular}
\end{table*}

\begin{figure*}[!ht]
    \centering
    \includegraphics[width=0.32\linewidth]{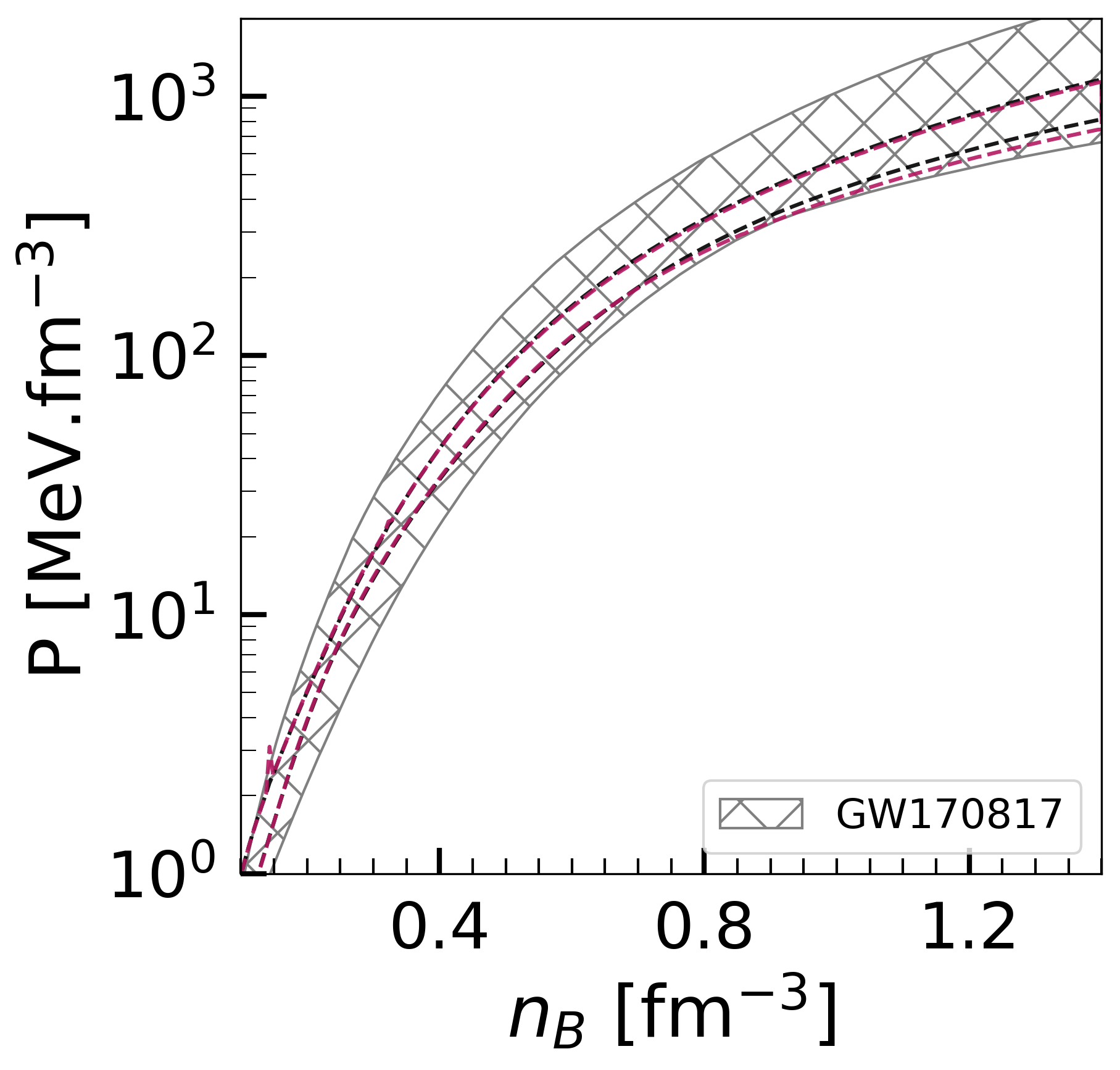} 
    \includegraphics[width=0.32\linewidth]{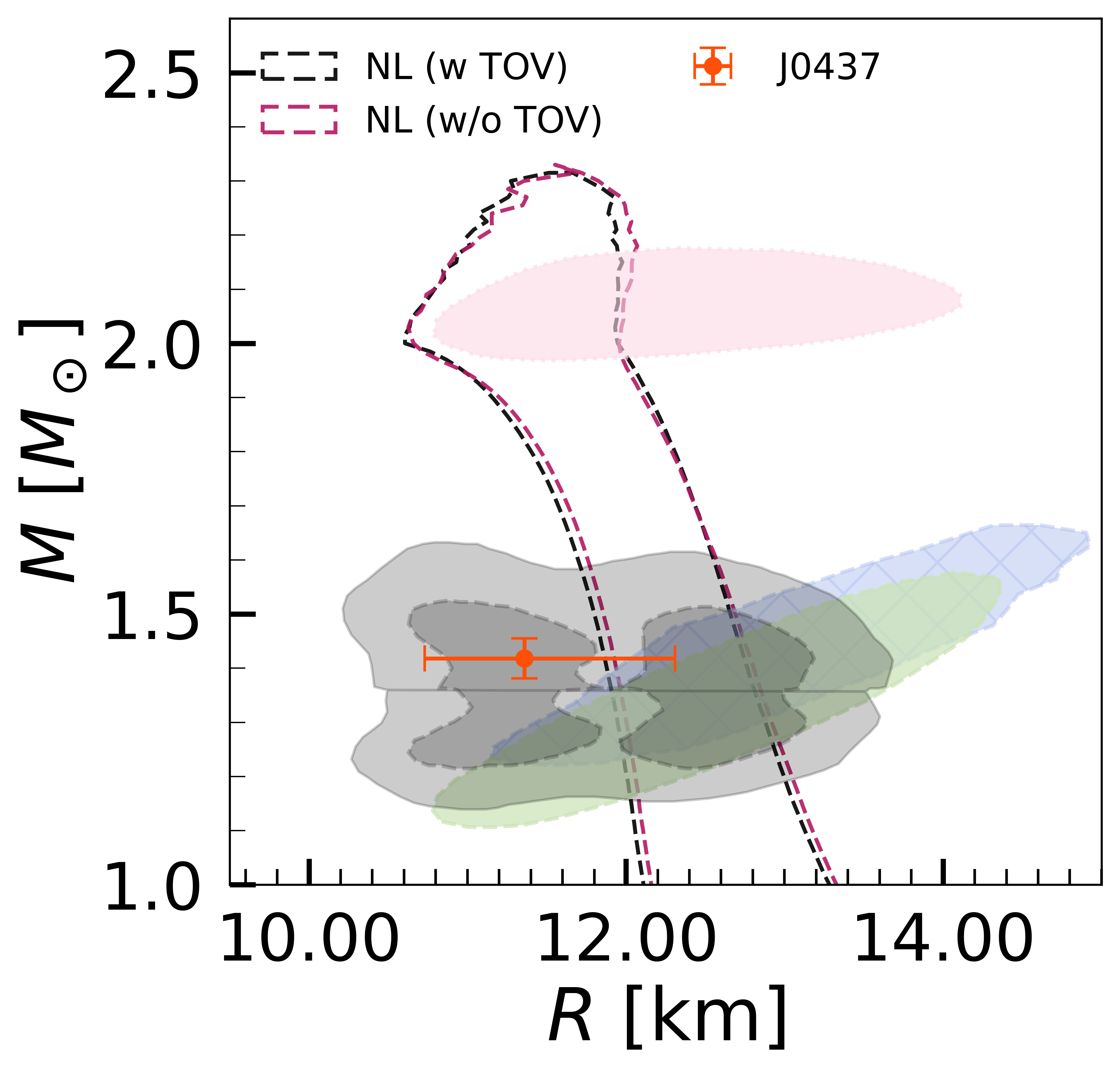} 
    \includegraphics[width=0.32\linewidth]{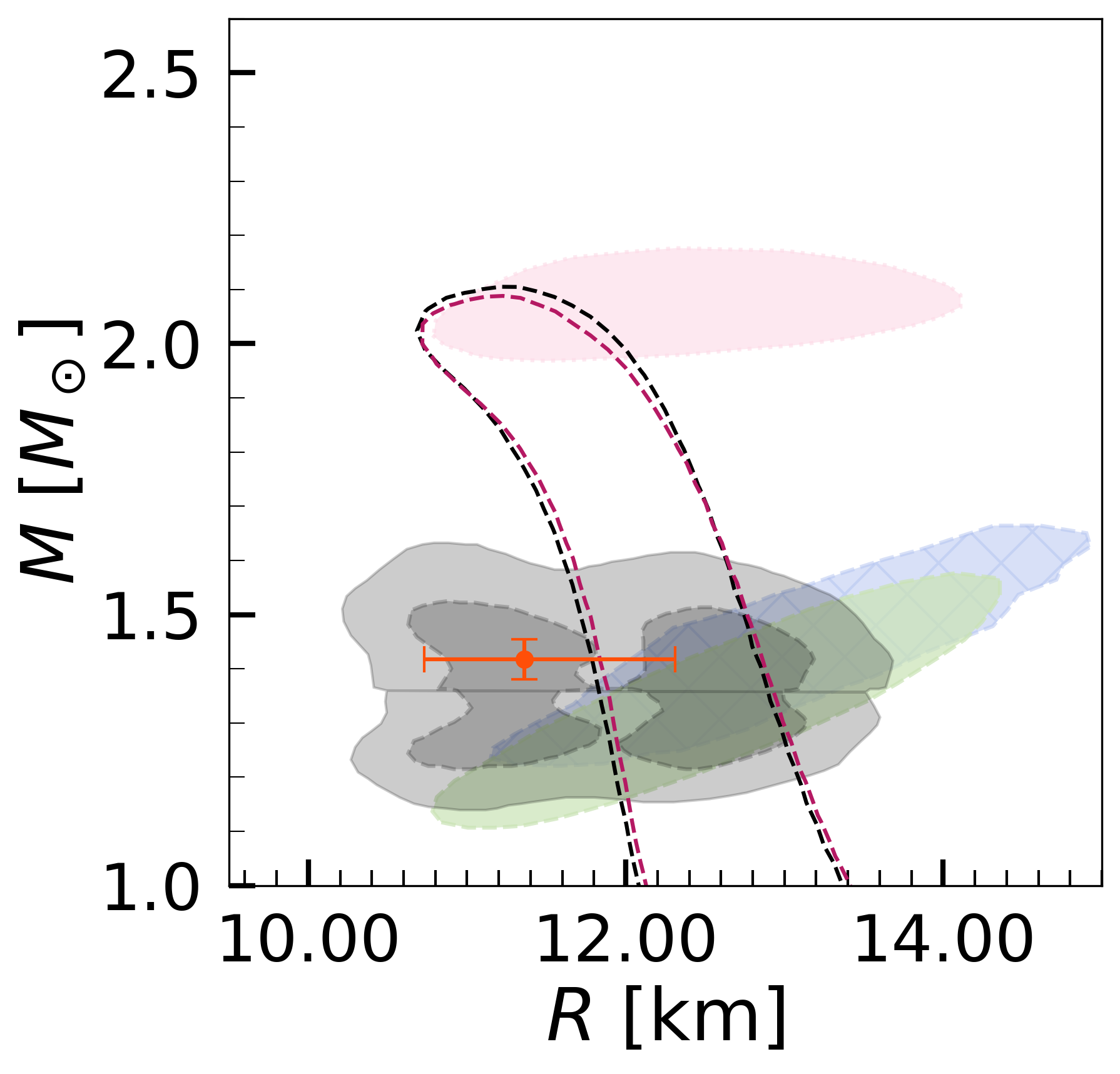}
    \caption{ The 90\% confidence interval (CI) for the EoS posterior pressure \(P\) as a function of the number density \(n_B\) is shown in the left panel, while the middle panel displays the model mass-radius posterior \(P(R|M)\). The right panel presents the two-dimensional mass-radius probability \(P(M, R)\). These results are derived from two different Bayesian analyses: (i) with the NS maximum mass obtained by solving the TOV equations (w TOV), and (ii) with the fitted  Eq.~(\ref{mmax}) (w/o TOV). For comparison, the left panel includes the 90\% CI for the equation of state inferred from GW170817 observations (grey shaded region), and the middle and right panels show constraints from the binary components of GW170817, with 90\% (solid) and 50\% (dashed) CIs respectively \cite{Abbott2018b}. The 68\% CI for the two-dimensional mass-radius posterior distribution of millisecond pulsars is also displayed: PSR J0030+0451 is shown in pastel blue and soft green \cite{Riley:2019yda, Miller:2019cac}, PSR J0740+6620 in blush pink \cite{Riley:2021pdl, Miller:2021qha}, and the latest NICER measurements for PSR J0437-4715 are shown in orange \cite{Choudhury:2024xbk}.}
    \label{fig4}
\end{figure*}

The constraints imposed on the equation of state within the Bayesian Inference framework, using astrophysical observations of neutron star properties, are quite time-consuming due to the need to solve the TOV equations. Particularly, solving the TOV equations for millions of parameterizations during Bayesian inference to ensure the EoS supports a NS maximum mass above 2 solar mass becomes very computational-intensive. Eq.~(\ref{mmax}) can be employed to circumvent the need to solve the TOV equations during parameter space sampling of an EoS model. 

We have conducted two identical Bayesian inferences using RMF models, {the models that define set NL}, incorporating various nuclear physics constraints — from neutron skin thickness to heavy ion collision (HIC) data, as discussed in Ref.~\cite{Tsang:2023vhh, Sorensen:2023zkk, Imam:2024xqg}. The key difference is that in one of the analyses we solve the TOV equations to determine NS maximum mass during sampling, and in the other, we use Eq.~(\ref{mmax}) instead. {Then the TOV equations need to be solved for both inferences, but only for the sampled EoS. } 
To determine the likelihood associated with the maximum NS mass, we use a Fermi-Dirac likelihood function to ensure that the maximum NS mass exceeds $2 M_\odot$, with a {short} tail to aid in sampling convergence, represented by the equation:
\bea
\mathcal{L}^{\rm NS}(D |\boldsymbol{\theta}) = \frac{1}{\exp \left(- {\frac{ m(\boldsymbol{\theta})-2.0}{0.01}} \right) + 1}. \label{likelihood}
\eea
We used the PyMultiNest sampler \cite{Buchner:2014nha, Buchner:2021kpm} with 1500 live points on the Deucalion high performance computing (HPC) system at the University of Minho, Portugal, using 4 nodes with 128 CPUs each. The summary of CPU times for both analyses is shown in Table~\ref{tab1}. It shows a dramatic reduction in sampling time for the scenario where the TOV equation was bypassed. {This resulted in a reduction in CPU time of more than a factor of seven.}  We then examined the posterior distributions for the NS mass-radius relationships for both cases. The comparison involved around 8,000 EoSs for each case, as can be seen from Table~\ref{tab1}. 

Figure~\ref{fig4} shows the results of these two different analyses. The left panel shows the posterior distributions of pressure \(P\) over baryon density \(n_B\). {The other two panels correspond to two different representations of the mass-radius posterior distributions: the middle panel shows the 90\% confidence interval of the mass radius conditional probability \(P(R|M)\), and the right panel shows the two-dimensional mass radius probability distribution  \(P(M, R)\)} obtained for two different scenarios. Several constraints are also shown, including those from GW170817 observations and NICER X-ray data on millisecond pulsars. The only marginal differences between the two different scenarios indicate that Eq. (\ref{mmax}) is quite reliable.

\subsubsection{ Mass-Radius-Tidal deformability} \label{mapping-eos-mr-ml}

\begin{figure}[!ht]
 \centering
 \includegraphics[width=9cm,height=8cm]{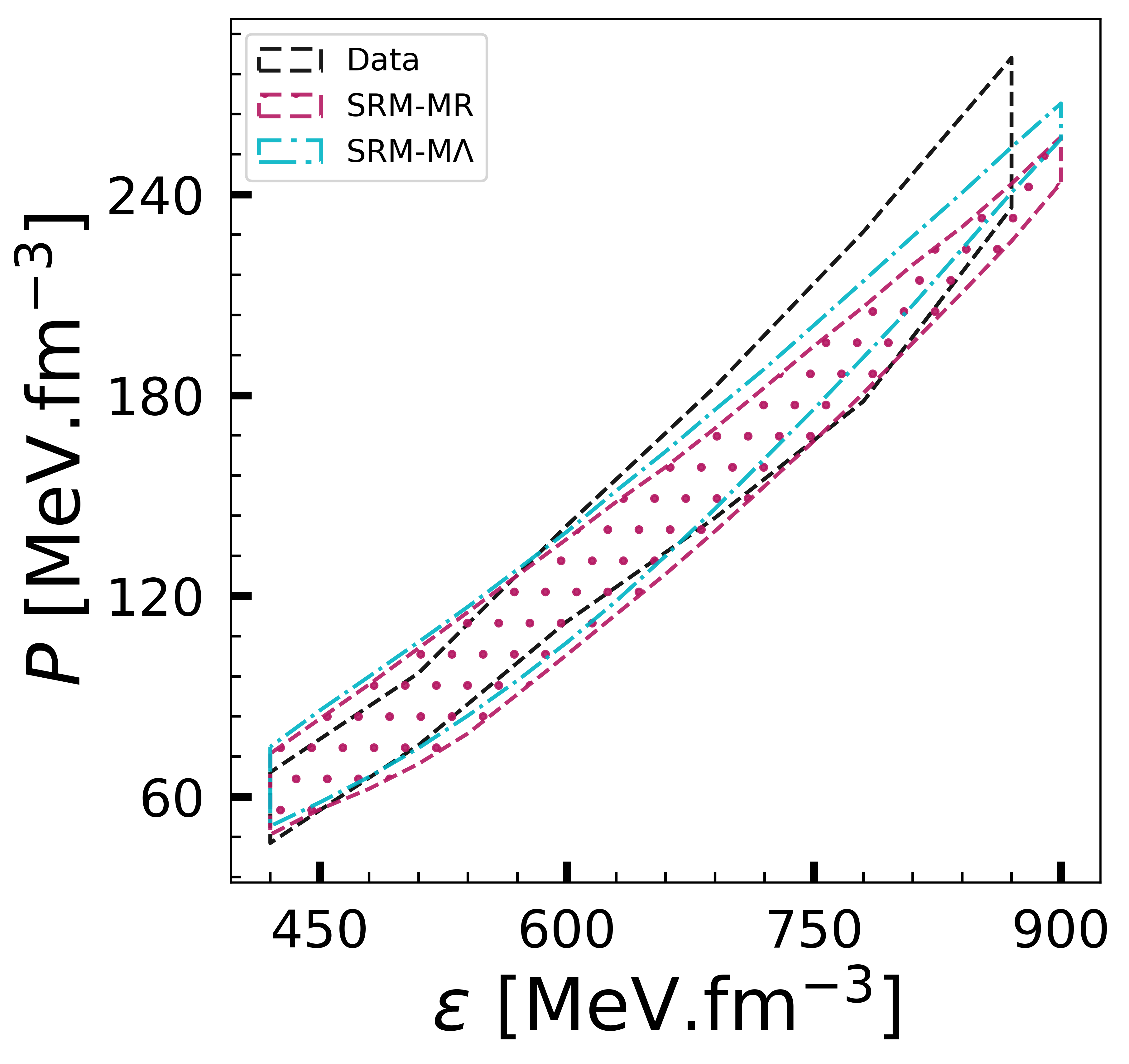}
 \caption{The 90\% confidence interval (CI) band of energy density and pressure for all 700 equations of state (black). The calculated 90\% CI bands for energy density and pressure are computed using Eqs. (\ref{eq-ec-mr}) \& (\ref{eq-pc-mr}) (light pink) and Eqs. (\ref{eq-ec-ml})\&(\ref{eq-pc-ml}) (cyan), as shown in the figure.}\label{fig5}
\end{figure} 

\begin{table*}[!ht]
\centering
\caption{The optimal equations obtained through feature and target variable minimization.  In this context, "{\bf Y}" specifically denotes the $\epsilon$, and  $P$, while the features "{\bf X}" pertain to the neutron star properties, i.e. mass $M$, radius $R$ and tidal deformability $\lambda$, in a mass range of 1-2 $M_\odot$. We provide the relative error values, ``RE" for all the datasets.}\label{tab2}
\setlength{\tabcolsep}{10.pt}
\renewcommand{\arraystretch}{1.4}
\begin{tabular}{cclcccccccc}
\hline \hline
 \multirow{2}{*}{\textbf{Y}}&\multirow{2}{*}{\textbf{X}} & \multirow{2}{*}{\textbf{Eqs.}}   & \multicolumn{8}{c}{\textbf{Datasets [RE (\%)]}} \\
\cline{4-11}
 &  & &  \textbf{NL} & \textbf{NL-hyp} &\textbf{CompOSE}& \textbf{CSE} & \textbf{PWP}& \textbf{DDB} & \textbf{SR} & \textbf{All}\\
\hline
\multirow{1}{*}{${\epsilon}$} & [M,R] &\multirow{1}{*}{\ref{eq-ec-mr}} & 3.4& 5.8 &6.2 &10.2 & 11.0 & 4.4 &4.5 &5.5\\
\multirow{1}{*}{${P}$} & [M,R] &\multirow{1}{*}{\ref{eq-pc-mr}} & 5.3& 5.3 &6.0 &7.8 & 9.3 & 5.2 &5.5 &5.7\\
\multirow{1}{*}{${\epsilon}$} &[M,$\lambda$] &\multirow{1}{*}{\ref{eq-ec-ml}} & 3.8& 7.0 &4.6 &7.4 & 7.5 & 4.0 &5.2 &5.1\\
\multirow{1}{*}{${P}$} & [M,$\lambda$] &\multirow{1}{*}{\ref{eq-pc-ml}}  & 6.5& 7.9 &8.4 &11.9 & 8.9 & 6.7 &7.4 &7.5\\
\hline \hline
\end{tabular}
\end{table*}

We use the symbolic regression method to express the central energy density and pressure of the neutron star in terms of bulk NS properties such as mass, radius and tidal deformability.  In particular, our \textbf{Y} vector comprises the energy density $\epsilon$ and pressure $P$ at the centre for mass range from 1.0 - 2.0 $M_\odot$ for each EoS, while the mass, radius and tidal deformability are stored in the vector \textbf{X}. We first consider the expression of $\epsilon$ and $P$ in terms of the NS mass ($M$) and radius ($R$). The optimal equations are obtained as

\bea
\frac{\epsilon}{\text{MeV} \cdot \text{fm}^{-3}} &=& \frac{15500 \left( \left( \frac{M}{M_\odot} \right)^3 + 1.018 \left( \frac{M}{M_\odot} \right)^2 + \frac{M}{M_\odot} \right)}
{\left( \frac{R}{\text{km}} \right)^2} \nonumber\\
&&- \frac{3100 \left( \frac{R}{\text{km}} \right) \left( \frac{M}{M_\odot} \right)^2+46747.711}{\left( \frac{R}{\text{km}} \right)^2 } \, , \label{eq-ec-mr}
\eea
\bea
\frac{P}{\text{MeV} \cdot \text{fm}^{-3}} 
&=&
\frac{5 \times 10^{5.540} \times 10^{\left( \frac{M}{M_\odot} \right)}}{\left( \frac{R}{\text{km}} \right)^{5.540}} + 18.553 \, . \label{eq-pc-mr}
\eea

Similarly, we have performed a symbolic regression to map the EoS parameters onto the NS mass and corresponding tidal deformability. We considered the dimension full tidal deformability $\lambda= (\frac{GM}{c^2})^5 \times \Lambda$, which explicitly depends on the NS mass. We obtained the optimal equations as
%
\bea
\frac{\epsilon}{\text{MeV} \cdot \text{fm}^{-3}}&=&
155\left( \left(\frac{M}{M_\odot}\right)^2\!\!\!\!+\left(\frac{M}{M_\odot}\right)\!+ 30.424\right)^{1.066\left(\frac{\lambda}{km^{5}}\right)^{-0.033}}\nonumber\\
&-&\!1920.053 \, , \label{eq-ec-ml}
\eea
\bea
\frac{P}{\text{MeV} \cdot \text{fm}^{-3}} = \frac{873161.216\left(\frac{M}{M_\odot}\right)}{\left(\frac{\lambda}{km^{5}}\right) + 2057.082} \, . \label{eq-pc-ml}
\eea

In Table~\ref{tab2} we summarize the RE (in \%) between actual and predicted values of $\epsilon$ and $P$, for different sets of EoSs considered. The small values of the RE give insight into the accuracy of these predictions. Note that when considering the pressure functional, the RE values are slightly higher than those for the energy density when considering the tidal deformability. This may be due to the larger coefficient values in the tidal deformability equations. For the RMF models within the NL, NL-hyp, DDB, SR and CompOSE families, the relative errors in the EoSs are moderately reduced, as shown in Table~\ref{tab2}, with the relative error generally not exceeding about $10$\%.
The relative error obtained by combining the results for all sets of EoS is about $5-8\%$.

\subsubsection*{Applications}

The equations of state can be immediately obtained, from the observed mass-radius-tidal deformability, by using separately Eqs.~(\ref{eq-ec-mr}) \& (\ref{eq-pc-mr})(SRM-MR) and Eqs.~(\ref{eq-ec-ml}) \& (\ref{eq-pc-ml})(SRM-M$\lambda$).
To test the reliability of these fits, we selected {from each of the different sets shown in}  Figure~\ref{fig2}, {100} MR and {100} M$\lambda$ curves to reconstruct the EoSs using the relationships in Eqs.~(\ref{eq-ec-mr}-\ref{eq-pc-ml}). These {700} reconstructed EoSs are compared to the original EoSs (All) in Figure~\ref{fig5}. The 90\% confidence interval for the variation of pressure with energy density reconstructed using SRM-MR (light pink) and SRM-M$\lambda$ (cyan) are compared to the original EoSs (black). 
The reconstructed bands show a good overlap with the original one {in the range $\epsilon\sim 450-700$ MeV/fm$^3$}.  It may be noted that the deviation for the reconstructed EoSs increases somewhat with density.  
We have also repeated the calculations for different sets of randomly selected EoSs, the results remained more-or-less the same.

\section{Conclusions}\label{summary}

We have investigated the connections between neutron star properties and their central energy density and pressure using seven diverse sets of equations of state.  
These EoSs correspond to a variety of models, including both agnostic frameworks and relativistic mean field models. The NSs were assumed to consist of nucleons, hyperons, and other exotic degrees of freedom in beta equilibrium. Our study shows that the maximum mass of a neutron star is strongly correlated with the pressure and the baryon density at an energy density of about 800 MeV fm$^{-3}$. Using symbolic regression methods, we are thus able to construct an analytical relationship capturing this dependence.
We also observed that the energy density and pressure of neutron stars are strongly connected to certain combinations of their mass, radius, and tidal deformability in the mass range 1-2 $M_\odot$.

We performed the Bayesian analysis to constrain the EoS by incorporating the relationship between the maximum mass of NS and pressure (Eq. (\ref{mmax})) along with empirical constraints from finite nuclei and heavy-ion collision experiments. The posterior distributions of EoSs and MR curves are nearly the same as those in which the maximum NS mass was calculated by solving the TOV equations. However, the Bayesian inference for the former case is faster by a factor of seven. 

We have obtained the expressions relating the equations of state to the NS observables such as the radius, and tidal deformability in the mass range of 1-2 $M_\odot$. The EoSs derived using these expressions (Eqs.~(\ref{eq-ec-mr}-\ref{eq-pc-ml})) using NS observables are found to have good overlap with the actual EoSs.

The results underscore the potential of combining symbolic regression and Bayesian inference to efficiently and accurately constrain the EoS of dense matter, offering valuable insights for the study of neutron star interiors and the properties of dense QCD matter.

\section*{Acknowledgements} 

We thank M. Oertel for providing us the EoS models from the CompOSE database.
This work was produced with the support of Deucalion HPC, Portugal, in the Advanced Computing Project 2024.14108.CPCA.A3,  RNCA (Rede Nacional de Computação Avançada), funded by the FCT (Fundação para a Ciência e a Tecnologia, IP, Portugal) with DOI identifier \href{https://sciproj.ptcris.pt/20234614108466780676546653PCA}{10.54499/2023.14108.CPCA.A3}.
NKP would like to acknowledge CFisUC, University of Coimbra, for their hospitality and local support provided during his visit for the purpose of conducting part of this research, and he would also like to thank the Department of Science and Technology, Ministry of Science and Technology, India, for the support of DST/INSPIRE Fellowship/2019/IF190058. NKP is also grateful to the Science and Engineering Research Board (SERB), Govt. of India, for the international travel support (ITS/2023/002096). KZ acknowledge support by the CUHK-Shenzhen University development fund under grant No. UDF01003041 and UDF03003041, and Shenzhen Peacock fund under No. 2023TC0179. This work was also partially supported by national funds from FCT (Fundação para a Ciência e a Tecnologia, I.P, Portugal) under projects 
UIDB/04564/2020 and UIDP/04564/2020, with DOI identifiers 10.54499/UIDB/04564/2020 and 10.54499/UIDP/04564/2020, respectively, and project 2022.06460.PTDC with the  DOI identifier 10.54499/2022.06460.PTDC. H.P. acknowledges the grant 2022.03966.CEECIND (FCT, Portugal) with DOI identifier 10.54499/2022.03966.CEECIND/CP1714/CT0004.




\end{document}